\documentclass[a4paper,11pt]{article}
\usepackage{pos}
\usepackage{dsfont}
\usepackage{url}
\usepackage{braket}
\usepackage{hyperref}

\newcommand{\e}{\mathrm{e}}
\newcommand{\im}{\mathrm{i}}
\newcommand{\diff}{\mathrm{d}}
\newcommand{\gTr}{\mathrm{gTr}}

\title{Implementation of bond weighting method for the Grassmann tensor renormalization group}
\ShortTitle{Implementation of bond weighting method for the GTRG}

\author*[a,b]{Shinichiro Akiyama}

\affiliation[a]{Center for Computational Sciences, University of Tsukuba,
Ibaraki, 305-8577, Japan}

\affiliation[b]{Endowed Chair for Quantum Software, University of Tokyo,
Tokyo, 113-0033, Japan}

\emailAdd{akiyama@ccs.tsukuba.ac.jp}

\abstract{
We demonstrate the efficiency of the bond weighting method for the Grassmann tensor renormalization group (TRG).
Benchmarking with the two-dimensional Gross-Neveu model with the Wilson fermion at finite density, we show that the bond weighting method improves the accuracy of the original Grassmann TRG.
We also provide a sample code of the bond-weighted TRG that can be applied to the two-dimensional models including fermions on a square lattice.
}

\notes{
\note{UTCCS-P-150}
}

\FullConference{The 40th International Symposium on Lattice Field Theory (Lattice 2023)\\
July 31st - August 4th, 2023\\
Fermi National Accelerator Laboratory\\}

\begin{document}
\maketitle

\section{Introduction}

The tensor renormalization group (TRG) is a typical type of tensor network method and can be regarded as a variant of the real-space renormalization group.
TRG approach provides us with a way to evaluate partition functions by representing them as tensor networks.
One of the important features of this approach is that there is no sign problem as in the standard Monte Carlo simulation since we do not resort to any probabilistic interpretation for the given Boltzmann weight.
Since Levin and Nave proposed a TRG algorithm for the two-dimensional classical system~\cite{Levin:2006jai}, the method has been extended in various ways.
One of the interesting extensions of the TRG approach is the so-called Grassmann TRG, which enables us to compute the fermionic path integrals without introducing any pseudo fermions~\cite{Gu:2010yh}.
In the high-energy community, the Grassmann TRG was first applied to the Schwinger model~\cite{Shimizu:2014uva}.
Since fermions obey Pauli's excursion rule, fermionic systems are well suited to be represented by tensor networks or qubits~\cite{Banuls:2019bmf,Funcke:2023jbq}.
In the case of the path-integral formalism, fermions are described by the Grassmann numbers and we can immediately express the given fermionic path integral as a Grassmann tensor network.
There is a systematic way to construct a Grassmann tensor network representation for a wide class of lattice fermion theories~\cite{Akiyama:2020sfo}.

In this study, we apply the bond weighting (BW) method recently proposed in Ref.~\cite{PhysRevB.105.L060402} for the Grassmann TRG.
The BW method is a novel way to improve the accuracy of the TRG.
Interestingly, the BW method allows us to improve the accuracy without increasing the computational complexity of the original TRG algorithm.
This is the most different aspect from the conventional techniques to improve the TRG such as the tensor network renormalization (TNR)~\cite{PhysRevLett.115.180405}, where we need to obtain the optimized disentanglers and isometric tensors.
\footnote{
Recently, the TNR-type algorithms have also been extended to the fermionic systems~\cite{Asaduzzaman:2022pnw}.
}

This paper is organized as follows. In section~\ref{sec:model}, we introduce the Grassmann tensor network representation of the two-dimensional Gross-Neveu model with the Wilson fermion as an example.
We briefly explain the BW method in section~\ref{sec:bwm}. 
Numerical results are shown in section~\ref{sec:results} and section~\ref{sec:summary} is devoted to summary.

\section{Grassmann tensor network representation for the Gross-Neveu model}
\label{sec:model}

We consider the single-flavor Gross-Neveu model with the Wilson fermion on a square lattice.
The model is defined by the following action,
\begin{align}
\label{eq:action}
	S
	&=-\frac{1}{2}
	\sum_{n\in\Lambda_{2}}\sum_{\nu=1,2}
	\left[
	\e^{\mu\delta_{\nu,2}}\bar{\psi}(n)(r\mathds{1}-\gamma_{\nu})\psi(n+\hat{\nu})
	+\e^{-\mu\delta_{\nu,2}}\bar{\psi}(n+\hat{\nu})(r\mathds{1}+\gamma_{\nu})\psi(n)
	\right]
	\nonumber\\
	&+(m+2r)\sum_{n}\bar{\psi}(n)\psi(n)
	-\frac{g^{2}}{2}\sum_{n}
	\left[(\bar{\psi}(n)\psi(n))^{2}
	+(\bar{\psi}(n)\im\gamma_{5}\psi(n))^{2}
	\right]
	,
\end{align}
where $r$, $m$, $g^{2}$, $\mu$ denote the Wilson parameter, mass, coupling constant, and chemical potential.
The Grassmann TRG study of this model was first made in Ref.~\cite{Takeda:2014vwa}.
$\psi(n)$ and $\bar{\psi}(n)$ are the two-component Grassmann-valued fields and $\gamma_{\nu}$'s are the two-dimensional $\gamma$-matrices.
We define the path integral as
\begin{align}
\label{eq:path_integral}
	Z
	=
	\int
	\left(
	\prod_{n}
	\diff\psi(n)
	\diff\bar{\psi}(n)
	\right)
	~\e^{-S}
	,
\end{align}
and the path integral measure is given by $\diff\psi(n)\diff\bar{\psi}(n)=\diff\psi_{1}(n)\diff\bar{\psi}_{1}(n)\diff\psi_{2}(n)\diff\bar{\psi}_{2}(n)$.
Following Ref.~\cite{Akiyama:2020sfo}, we can immediately derive the Grassmann tensor network representation for Eq.~\eqref{eq:path_integral} as
\begin{align}
\label{eq:gtn}
	Z
	=
	\gTr\left[
		\prod_{n}\mathcal{T}_{\eta(n)\zeta(n)\bar{\eta}(n-\hat{1})\bar{\zeta}(n-\hat{2})}
	\right]
	,
\end{align}
where $\mathcal{T}$ is a Grassmann tensor defined on each lattice site and $\eta(n)$ and $\bar{\eta}(n)$ are the auxiliary two-component Grassmann fields living on the spatial link $(n,\hat{1})$. 
Similarly, $\zeta(n)$ and $\bar{\zeta}(n)$ are living on the temporal link $(n,\hat{2})$. 
$\gTr$ in the right-hand side of Eq.~\eqref{eq:gtn} is a short-hand notation of $\int\prod_{n}\diff\bar{\eta}(n)\diff\eta(n)\diff\bar{\zeta}(n)\diff\zeta(n)\e^{-\bar{\eta}(n)\eta(n)-\bar{\zeta}(n)\zeta(n)}$.
Since Eq.~\eqref{eq:action} is translationally invariant on a lattice, the resulting Grassmann tensor network in Eq.~\eqref{eq:gtn} is uniform.
Hereafter, we omit the site dependence in the auxiliary Grassmann fields.
The Grassmann tensor $\mathcal{T}$ in Eq.~\eqref{eq:gtn} is defined as
\begin{align}
	\mathcal{T}_{\eta\zeta\bar{\eta}\bar{\zeta}}
	=
	\sum_{i_{1},j_{1}}\sum_{i_{2},j_{2}}\sum_{i'_{1},j'_{1}}\sum_{i'_{2},j'_{2}}
	T_{i_{1}j_{1}i_{2}j_{2}i'_{1}j'_{1}i'_{2}j'_{2}}
	\eta_{1}^{i_{1}}
	\zeta_{1}^{j_{1}}
	\eta_{2}^{i_{2}}
	\zeta_{2}^{j_{2}}
	\bar{\zeta}_{1}^{j'_{1}}
	\bar{\eta}_{1}^{i'_{1}}
	\bar{\zeta}_{2}^{j'_{2}}
	\bar{\eta}_{2}^{i'_{2}}
	.
\end{align}
$T_{i_{1}j_{1}i_{2}j_{2}i'_{1}j'_{1}i'_{2}j'_{2}}$ is a normal tensor whose index is given by an eight-bit string.
If we use the chiral representation such that $\gamma_{1}=\sigma_{x}$, $\gamma_{2}=\sigma_{y}$, and $\gamma_{5}=-\im\gamma_{1}\gamma_{2}=\sigma_{z}$, one finds
\begin{align}
\label{eq:c_tensor}
	&T_{i_{1}j_{1}i_{2}j_{2}i'_{1}j'_{1}i'_{2}j'_{2}}
	=
	(-1)^{P_{i_{1}j_{1}i_{2}j_{2}i'_{1}j'_{1}i'_{2}j'_{2}}}
	\exp\left[\frac{\mu}{2}(i_{2}-j_{2}+i'_{2}-j'_{2})\right]
	\left(\frac{1}{\sqrt{2}}\right)^{i_{1}+j_{1}+i_{2}+j_{2}+i'_{1}+j'_{1}+i'_{2}+j'_{2}}
	\nonumber\\
	&\times
	\left[
	\left((m+2r)^{2}+2g^{2}\right)
	\delta_{i_{1}+i_{2}+j'_{1}+j'_{2},0}
	\delta_{j_{1}+j_{2}+i'_{1}+i'_{2},0}
	-(m+2r)
	\delta_{i_{1}+i_{2}+j'_{1}+j'_{2},1}
	\delta_{j_{1}+j_{2}+i'_{1}+i'_{2},1}
	\right.
	\nonumber\\
	&
	\left.
	-(-1)^{i_{1}+i_{2}+j_{2}+i'_{1}}
	(+\im)^{i_{2}+j_{2}+i'_{2}+j'_{2}}
	(m+2r)
	\delta_{i_{1}+i_{2}+j'_{1}+j'_{2},1}
	\delta_{j_{1}+j_{2}+i'_{1}+i'_{2},1}
	-
	\bar{A}_{i_{1}i_{2}j'_{1}j'_{2}}
	A_{j_{1}j_{2}i'_{1}i'_{2}}
	\right]
	,
\end{align}
where we have introduced several tensors for convenience such that
\begin{align}
\label{eq:p_tensor}
	P_{i_{1}j_{1}i_{2}j_{2}i'_{1}j'_{1}i'_{2}j'_{2}}
	&=i_{1}(j_{1}+j_{2}+i'_{1}+i'_{2})
	+i_{2}(j_{2}+i'_{1}+i'_{2})
	+j'_{1}(i'_{1}+i'_{2})
	+j'_{2}i'_{2}
	+i'_{1}+i'_{2}
	,
\end{align}
\begin{align}
\label{eq:bar_tensor}
	\bar{A}_{i_{1}i_{2}j'_{1}j'_{2}}
	=
	\int\int\diff\bar{\psi}_{2}\diff\bar{\psi}_{1}
	(\bar{\psi}_{1}+\im\bar{\psi}_{2})^{j'_{2}}
	(\bar{\psi}_{1}+\bar{\psi}_{2})^{j'_{1}}
	(\bar{\psi}_{1}-\im\bar{\psi}_{2})^{i_{2}}
	(\bar{\psi}_{1}-\bar{\psi}_{2})^{i_{1}}
	,
\end{align}
\begin{align}
\label{eq:a_tensor}
	A_{j_{1}j_{2}i'_{1}i'_{2}}
	=
	\int\int\diff\psi_{2}\diff\psi_{1}
	(\psi_{1}+\im\psi_{2})^{i'_{2}}
	(\psi_{1}-\psi_{2})^{i'_{1}}
	(\psi_{1}-\im\psi_{2})^{j_{2}}
	(\psi_{1}+\psi_{2})^{j_{1}}
	.
\end{align}
$T_{i_{1}j_{1}i_{2}j_{2}i'_{1}j'_{1}i'_{2}j'_{2}}$ is implemented in our sample code~\cite{gbtrg}. 
The code assumes $r=1$ and the other parameters can be chosen to any value.
See ~${\texttt{module\_initial\_tensor.f90}}$ in Ref.~\cite{gbtrg}.
\footnote{
In the practical TRG computation, it is useful to map a two-bit-string index, say $(i_{1}j_{1})$, to a single index $x\in\mathds{Z}_{4}$.
This mapping allows us to regard $T_{i_{1}j_{1}i_{2}j_{2}i'_{1}j'_{1}i'_{2}j'_{2}}$ in Eq.~\eqref{eq:c_tensor} as a four-leg tensor $T_{xtx't'}$, which is ready to be contracted by TRG algorithms.
See Ref.~\cite{Akiyama:2021nhe} for a more detailed explanation.
}

\section{Bond weighting method}
\label{sec:bwm}

\begin{figure}[htbp]
	\centering
	\vskip10mm
	\hspace{-25truemm}
   	\includegraphics[width=0.5\hsize,bb=0 0 550 442]{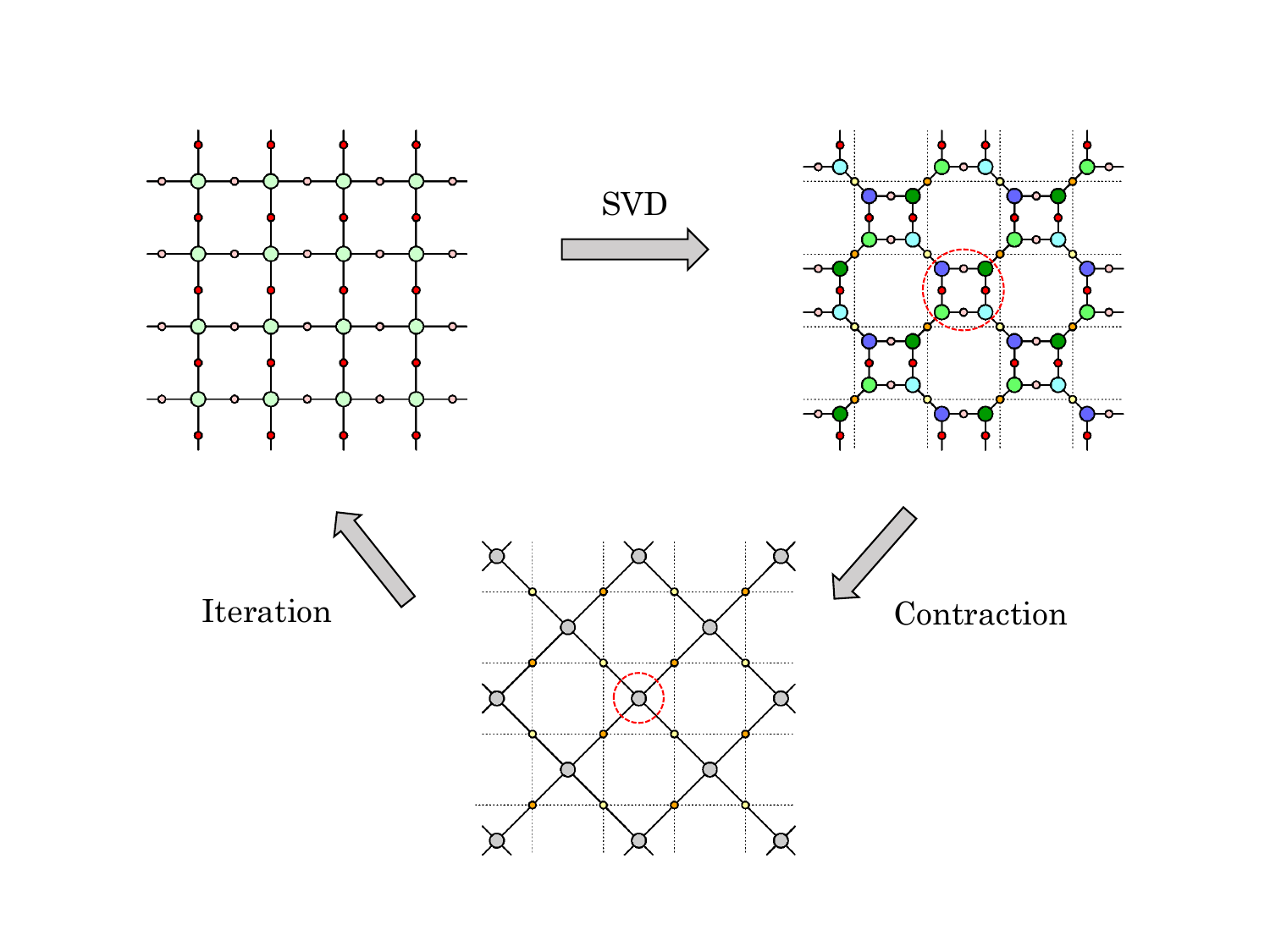}
	\vskip2mm
  	\caption{Schematic picture of the TRG with the BW method on a square tensor network.}
  	\label{fig:gbtrg}
\end{figure}

The BW method gives us a kind of generalization of the Levin-Nave TRG, which uses the singular value decomposition (SVD) to split each four-leg tensor on a lattice site into two three-leg tensors.
The BW method modifies it in the following way;
\begin{align}
\label{eq:bw_svd}
	T_{ab}
	\approx
	\sum_{c=1}^{D}
	U_{ac}
	\sigma_{c}^{(1-k)/2}
	\sigma_{c}^{k}
	\sigma_{c}^{(1-k)/2}
	V^{*}_{bc}
	,
\end{align}
where $D$ is the bond dimension and $k$ is a real-valued hyperparameter.
When we set $k=0$, Eq.~\eqref{eq:bw_svd} is nothing but the decomposition employed in the Levin-Nave TRG.
With the non-zero $k$, $\sigma^{k}$ in Eq.~\eqref{eq:bw_svd} can be regarded as some weight matrix defined on each edge in the tensor network.
After the decomposition in Eq.~\eqref{eq:bw_svd}, we can define a coarse-grained tensor by contracting four kinds of three-leg tensors and four weight matrices.
The procedure of bond-weighted TRG is schematically given in Fig.~\ref{fig:gbtrg}.
According to the discussion in Ref.~\cite{PhysRevB.105.L060402}, the optimal choice of $k$ is determined by the power counting for the singular value of the local tensor.
By the bond-weighted TRG, the coarse-grained local tensor on a square lattice is characterized by $(\sigma^{k})^4(\sigma^{(1-k)/2})^4=\sigma^{2k+2}$.
The SVD of this coarse-grained tensor gives a new singular value $\sigma'$.
Assuming that the coarse-grained tensor is characterized just by the singular value, not affected by unitary matrices, whose spectrum becomes scale invariant after sufficient times of TRG transformations, we can equate $\sigma^{2k+2}=\sigma$, which results in $k=-1/2$.
It is numerically confirmed that $k=-1/2$ is the optimal choice benchmarking with the ferromagnetic classical Ising model~\cite{PhysRevB.105.L060402} and massless free Wilson fermion~\cite{Akiyama:2022pse}.

The bond-weighted TRG for Grassmann path integrals is also implemented in our sample code~\cite{gbtrg}. 
You can choose the bond dimension $D$, hyperparameter $k$, and the number of coarse-grainings $N$. 
The sample code gives you $\ln Z/V$, where the lattice volume $V$ is equal to $2^{N}$.
The numerical results are saved in ~${\texttt{results.dat}}$~.
All codes except ~${\texttt{module\_initial\_tensor.f90}}$~ and ~${\texttt{load\_input\_trg}}$~ in ~${\texttt{module\_setup.f90}}$~ are independent of initial tensor details.
Our sample code can be applied to any two-dimensional model on a square lattice just by modifying the initial tensor.
\footnote{
If you set ~${\texttt{phase = 0}}$~ in ~${\texttt{module\_phase.f90}}$~, the code is reduced to be the normal bond-weighted TRG accompanied with $\mathds{Z}_{2}$ symmetry.
}

\section{Results}
\label{sec:results}

\begin{figure}[htbp]
	\centering
	\vskip5mm
   	\includegraphics[width=0.55\hsize]{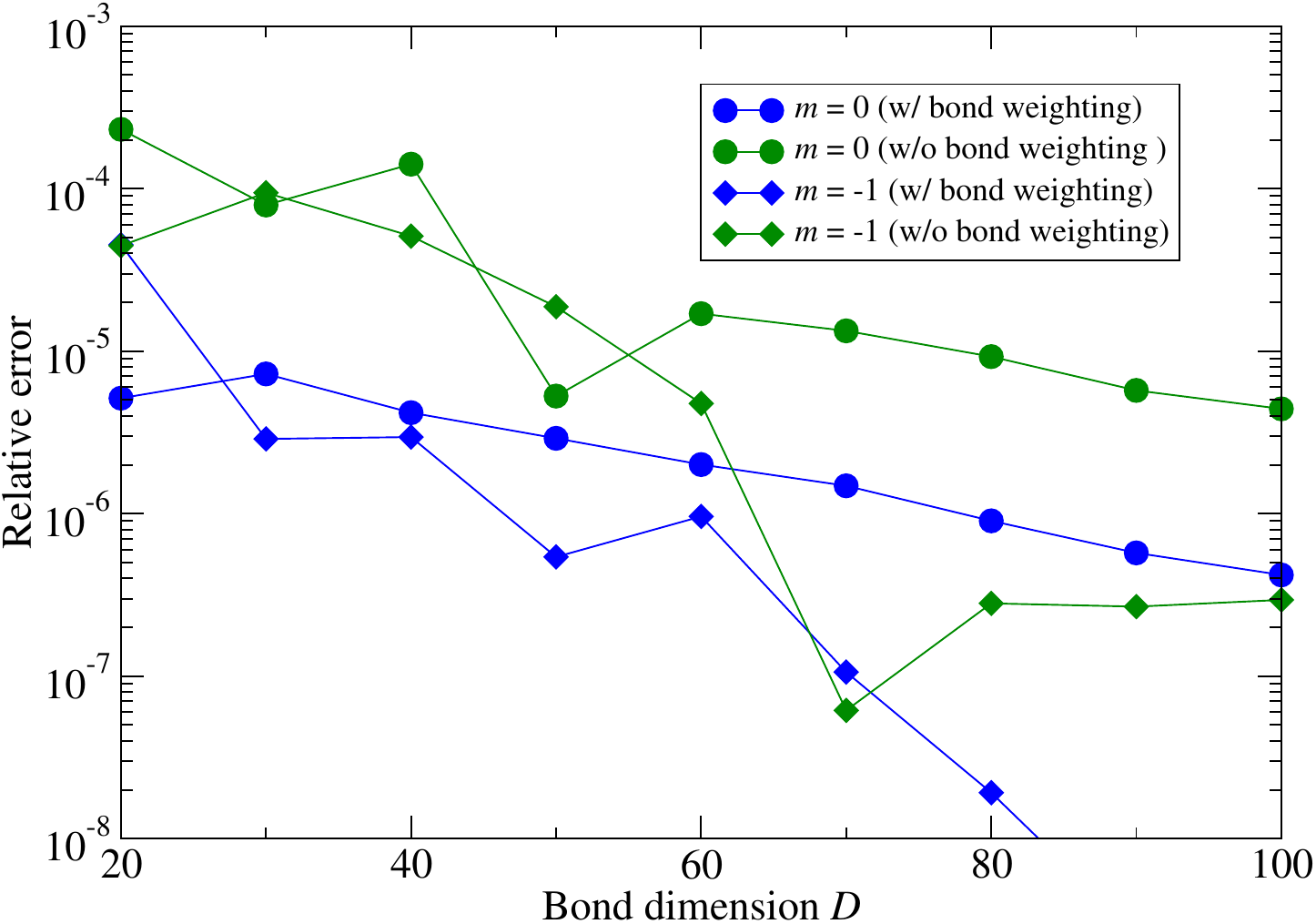}
	\vskip5mm
  	\caption{
	Relative error of the free energy as functions of bond dimension. 
	Blue (Green) symbols are obtained by the TRG with (without) the BW method.
	Circle and diamond symbols correspond to $m=0$ and $m=-1$, respectively.
	}
	\vskip5mm
  	\label{fig:comparison}
\end{figure}

We show several benchmark results obtained by our code~\cite{gbtrg}.
Firstly, we consider the free field model at $\mu=0$ which allows us to evaluate the relative error of the TRG calculation from the exact value.
Fig.~\ref{fig:comparison} compares the relative error of the free energy, $|\ln Z_{\rm exact}-\ln Z(D)|/|\ln Z_{\rm exact}|$, calculated by the TRG with and without the BW method.
For demonstration, we select two cases with $m=0$ and $m=-1$.
Since we set $r=1$, $m=0$ corresponds to one of the critical points with the parity symmetry breaking~\cite{Aoki:1983qi}.
The accuracy at $m=0$ is improved by an order of magnitude.
At $m=-1$, where the system is off-critical, the BW method still outperforms the original TRG particularly for the larger bond-dimension regime: the bond-weighted TRG achieves the $\mathcal{O}(10^{-9})$ accuracy at $D=90$, although the accuracy of the normal TRG is saturating around $\mathcal{O}(10^{-7})$ with $D\ge80$.

\begin{figure}[htbp]
	\begin{minipage}[htbp]{0.48\hsize}
		\centering
		\vskip2mm
    		\includegraphics[width=0.8\hsize]{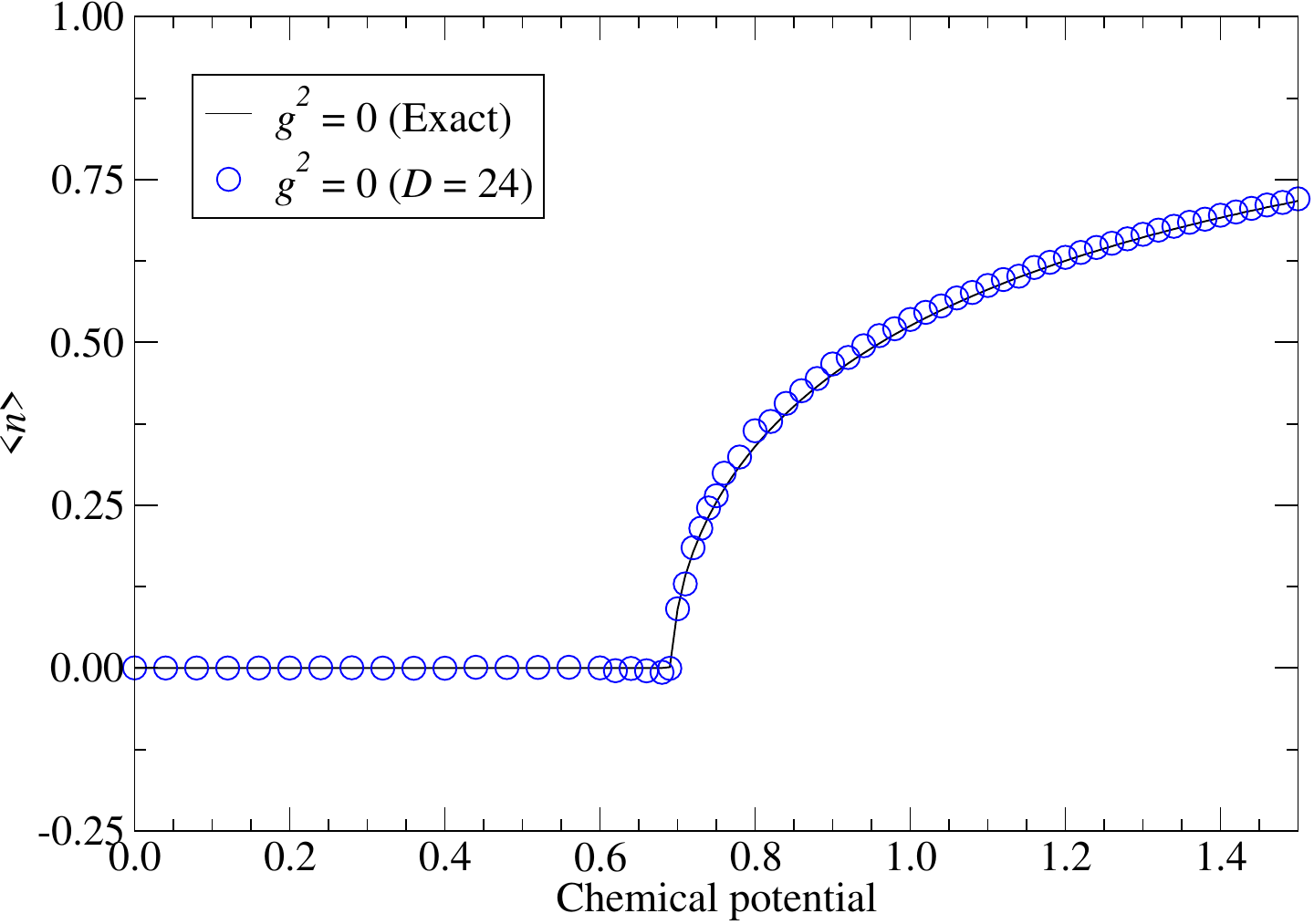}
		\vskip2mm
  		\caption{
		Number density as a function of $\mu$ at $m=-1$. 
		The solid curve is the exact result on the lattice with $V=2^{20}$. 
		Blue symbols show the bond-weighted TRG result with $g^{2}=0$.
		}
  		\label{fig:number_m_n1}
  	\end{minipage}
  	\hspace*{3mm}
	\begin{minipage}[htbp]{0.48\hsize}
    		\centering
		\vskip2mm
   	 	\includegraphics[width=0.8\hsize]{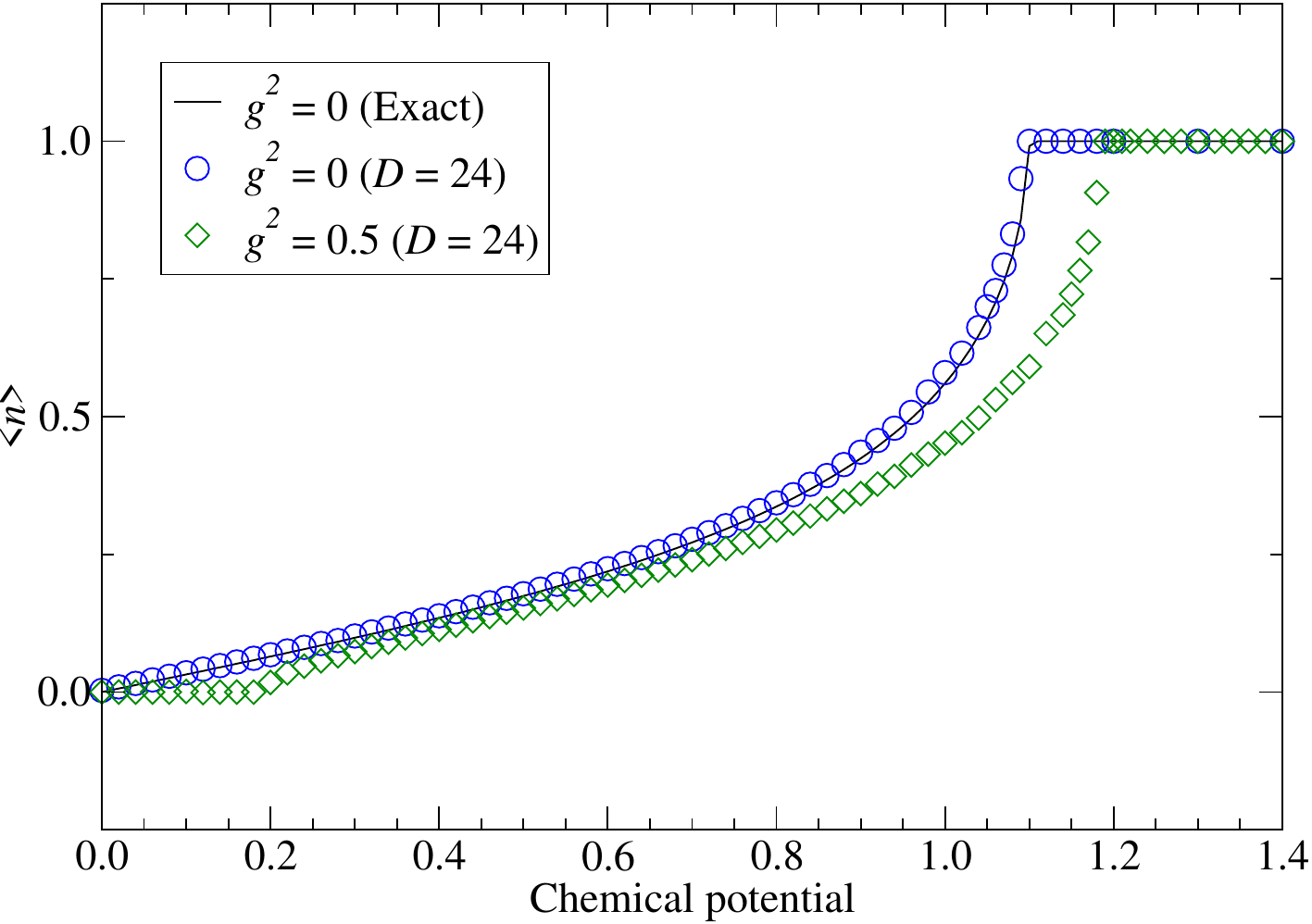}
		\vskip2mm
  		\caption{
		Number density as a function of $\mu$ at $m=0$. 
		The solid curve is the exact result on the lattice with $V=2^{20}$. 
		Blue (Green) symbols show the bond-weighted TRG result with $g^{2}=0$~($g^{2}=0.5$).
		}
  		\label{fig:number_m_0}
	\end{minipage}
\end{figure}

Fig.~\ref{fig:number_m_n1} shows the number density as a function of $\mu$ setting $(m,g^{2})=(-1,0)$. 
In the bond-weighted TRG computation, the number density is evaluated by the numerical difference via
\begin{align}
	\braket{n}
	=
	\frac{1}{V}\frac{\ln Z(\mu+\Delta\mu)-\ln Z(\mu)}{\Delta\mu}
	.
\end{align}
The resulting $\braket{n}$ by the bond-weighted TRG is in agreement with the exact one even with $D=24$.
We found that the numerical difference became unstable when we applied the TRG without BW at $D=24$.
Fig.~\ref{fig:number_m_0} shows the number density at $(m,g^{2})=(0,0)$ and $(0,0.5)$. 
Since $m=0$ corresponds to the critical point at $g^{2}=0$, there is no plateau of $\braket{n}=0$.
However, once we introduce finite $g^{2}$, $m=0$ is no longer the critical point and there should be the plateau of $\braket{n}=0$.
As a representative case, we set $g^{2}=0.5$ and the plateau of $\braket{n}=0$ is observed by the bond-weighted TRG.
We also confirmed that when we introduced a relatively small negative mass at $g^{2}=0.5$, the plateau of $\braket{n}=0$ shrunk.
This is in agreement with the weak-coupling expansion provided in Ref.~\cite{Kenna:2001na}: there is a critical line $m=-1.54g^{4}$ in the vicinity of $(m,g^{2})=(0,0)$.

\section{Summary}
\label{sec:summary}

We have confirmed that the BW method improves the accuracy of the TRG for fermionic systems.
As a demonstration, we have dealt with the Gross-Neveu model with the Wilson fermion on a square lattice.
The free field computation shows that the bond-weighted TRG achieves higher accuracy than the normal TRG.
Number densities computed by the numerical difference are consistent with the exact ones, even with $D=24$, noting that the initial tensor is characterized by $D=4$.
Our sample code is available on Ref.~\cite{gbtrg}, by which one can reproduce all the results provided in this paper.
One can also use the same code to compute other two-dimensional models on a square lattice just modifying the initial tensor.

\acknowledgments
We acknowledge the support from the Endowed Project for Quantum Software Research and Education, the University of Tokyo (\url{https://qsw.phys.s.u-tokyo.ac.jp/}) and JSPS KAKENHI Grant Number JP23K13096.
A part of the numerical calculation for the present work was carried out with ohtaka provided by the Institute for Solid State Physics, the University of Tokyo.


\bibliographystyle{JHEP}
\bibliography{bib/algorithm,bib/review,bib/formulation,bib/grassmann,bib/for_this_paper}

\end{document}